\documentclass[amsmath,amssymb,twocolumn]{revtex4-2}
\usepackage{graphicx}
\usepackage{subfig}
\usepackage{epsfig}
\usepackage{dcolumn}
\usepackage{multirow}
\usepackage{rotating}
\usepackage{verbatim}
\usepackage{bm}
\usepackage{color}
\usepackage{soul}
\usepackage{float}
\usepackage{hyperref}
\usepackage[utf8]{inputenc}

\def\lsim{\raise0.3ex\hbox{$<$\kern-0.75em\raise-1.1ex\hbox{$\sim$}}}
\def\gsim{\raise0.3ex\hbox{$>$\kern-0.75em\raise-1.1ex\hbox{$\sim$}}}

\def\beqa{\begin{eqnarray}}
\def\eeqa{\end{eqnarray}}

\begin{document}

\title{Multiplicity dependence of the $p_T$-spectra for charged particles and its relationship with partonic entropy}
\author{L.S. Moriggi$^{1}$}
\email{lucasmoriggi@unicentro.br}
\affiliation{$^{1}$ Universidade Estadual do Centro-Oeste (UNICENTRO), Campus Cedeteg, Guarapuava 85015-430, Brazil}
\author{G.S. Ramos $^{2}$} 
\email{silveira.ramos@ufrgs.br}
\author{M.V.T. Machado$^{2}$}
\email{magnus@if.ufrgs.br}
\affiliation{$^{2}$ High Energy Physics Phenomenology Group, GFPAE. Institute of Physics, Federal University of Rio Grande do Sul (UFRGS)\\
Caixa Postal 15051, CEP 91501-970, Porto Alegre, RS, Brazil} 

\begin{abstract}

We investigate the multiplicity dependence of the transverse momentum $p_T$ spectra of hadrons produced in high-energy collisions. We propose that the partonic distribution be parameterized by its non-extensive entropy and the parton saturation scale $Q_s(x)$. These two variables can be identified from the produced charged hadron distributions and provide important information on the gluon dynamics at the moment of interaction. From this perspective we interpret data from different ALICE multiplicity classes at $\sqrt{s}= 13$ TeV and $\sqrt{s}= 5.02$ TeV. A multiplicity dependent scaling function is presented  and the dependence of the interaction area on
multiplicity is also investigated.
\end{abstract}

\maketitle


\section{Introduction}

The transverse momentum, $p_T$, spectra are traditionally one of the main ways to obtain information about the dynamics of partons in the initial state of the interaction in proton-proton ($pp$), proton-nucleus ($pA$) and nucleus-nucleus ($AA$) collisions at high energies. The momentum distribution of the particles produced is sensitive to the geometric parameters of the collision, such as the average area of interaction,  $\langle A_{T}\rangle$, the nature of the projectile, the collision energy, $\sqrt{s}$, the pseudorapidity of the hadron produced, $\eta$, and the observed multiplicity class. Modifications of these collision parameters significantly change the shapes of the spectra, mainly in the region of large $p_T$, usually characterized by a power-law multiplicity $dN/d^2p_Td\eta \sim {p_{T}^{-m}}$ \cite {Arleo:2009ch,Brodsky:2005fza,Moriggi:2020zbv}. Mapping these observable modifications in terms of QCD degrees of freedom, quarks and gluons, requires a phenomenological analysis that allows connection between variables associated with partonic dynamics with the quantities that characterize the hadronic spectrum.

In \cite{Moriggi:2020zbv} we propose a power-law partonic transverse momentum distribution (TMD) function that essentially depends on two quantities: the power index $\delta n $ and the saturation scale $Q_s(x)$, where $x \sim p_T/\sqrt{s}$ is the fraction of the gluon's longitudinal momentum at central rapidity. These two quantities are easily identified in the final spectra of produced hadrons: the saturation scale is evident from the scaling with respect to the hard scale proven in the interaction $Q^2\sim p_T^2$ in the scaling variable, $\tau=Q^2/Q_s ^2(x)$, while the power index can be inferred from the slope of spectra at  the region of large $p_{T}$. The data description is relatively good by using these two quantities. In this work we intend to give a better justification of why this type of model works and its interpretation. In the context of non-extensive statistical mechanics, the power parameter is related to the entropic index $q$ of the Tsallis entropy \cite{Tsallis:1987eu,Tsallis:1998ws}. While in approaches based on Hawking-Unruh radiation in QCD applied to thermal hadronization \cite{Kharzeev:2005iz,Castorina:2007eb} the saturation scale plays the role of a temperature, $T=Q_s/(2\pi)$, and is sufficient to describe the system, in our model we also need to specify the power index (entropy). Thus, an analysis of the hadron spectrum on $p_{T}$ will provide us with information about partonic dynamics through its entropy $S_q$ and saturation scale $Q_s(x)$, with both quantities being necessary to characterize the system.

In \cite{Moriggi:2020zbv} we have already made an extensive study of the spectra of identified particle  based on their scaling properties in relation to the collision energy $\sqrt{s}$ at the energies of RHIC, TEVATRON and  LHC up to $\sqrt{s}= 13$ TeV. The underlying QCD dynamics is based on the high energy factorization or $k_T$-factorization \cite{Catani:1990xk,Catani:1990eg} where the building blocks are the transverse momentum distribution of partons (parton TMDs). The corresponding gluon TMD as function of gluon transverse momentum is denoted by $\phi (x,k_T)$. In this article we will investigate the behavior of the spectrum at a fixed energy for different multiplicity classes as defined by ALICE \cite{ALICE:2019dfi} in proton-proton  collisions at $\sqrt{s}= 5.02$ and 13 TeV. The $pp$ collisions at high multiplicity have shown properties close to those observed in nuclear collisions, which leads to the question about the relationship of initial/final state effects in these different collision systems. In our approach we intend to show that all the characteristics of the spectrum and its multiplicities can be well explained just by taking into account the properties of the initial state and the partonic entropy of the gluon system that take part in the initial interaction. The relationship between average transverse momentum and multiplicity of charged particles produced in different collision systems as $pp$, $pA$ and $AA$ \cite{ALICE:2013rdo} reactions presents a challenge to traditional models of particle production, being a fundamental indicator in order to distinguish effects of initial state and hydrodynamic evolution of the final state.

The main ideas underpinning the present study are as follows. The high energy collision probes a system of gluons with a probability distribution  $P(x,k_T)$ given by the Fourier transform to the QCD color dipole scattering amplitude, $\mathcal{N}(x,r)$, with $r$ being the transverse size of a color dipole. Due to the diffusion of gluons in momentum space given in a time scale $t\sim 1/x$, the variation of the distribution $P(x,k_T)$ is related to the probe of different substructures of the target. This process is described by an anomalous diffusion. We argue  that  in case the probability distribution $P(x,k_T)$ to be given by the entropy maximization two situations can occur: i) the Boltzmann-Gibbs entropy which in steady state results in a Gaussian distribution in transverse momentum; ii) the Tsallis entropy that in steady state results in our model containing a power (entropic) index. By maximizing a Tsallis entropy one associates a Lagrange multiplier $\beta$ to the average value of gluon transverse momentum $k_T$. Making use of scaling property very common to systems presenting anomalous diffusion one obtains $\langle k_T^2 \rangle \sim \beta^{-1} (x_s/x)^{1/3}$. This can be interpreted as a generalization of the Einstein relation and thus $\beta$ can be understood as the inverse of temperature. More specifically, we intend to explore the connection between the multiplicity of  produced hadrons, $dN/d^2p_{T_h}d\eta$, and partonic dynamics at high energies. In order to investigate such connection we define an indicator of partonic entropy associated with the diffusion of gluons in the $k_T$ space. Considering that in the equilibrium situation $\sqrt{s} \rightarrow \infty $ a behavior like $\propto k_T^{-4}$  is expected, in contrast to the Boltzmann-like exponential form, we consider that the most appropriate indicator is that of Tsallis \cite{Tsallis:1987eu,Tsallis:1998ws}, where the monotonic shape $\sim k_T^{-4}$ corresponds to the entropic index $q=3/2$. From this indicator, we observed a relationship between the growth of entropy, the area of interaction and the final multiplicity of charged hadrons.

This work is organized as follows. In section \ref{sec:modelA} the $k_T$-factorization formalism applied to description of  transverse momentum spectra of produced hadrons is presented. In section \ref{sec:modelB} we propose a partonic entropy indicator based on the formalism of non-extensive statistical mechanics and its implications for the production of hadrons at high energies. Main results are presented in section \ref{sec:results} where the model is compared with ALICE data for different multiplicities. Finally, in Sec. \ref{sec:conclusion}, the  main conclusions are summarized.

\begin{figure*}[t]
\includegraphics[width=0.8\textwidth]{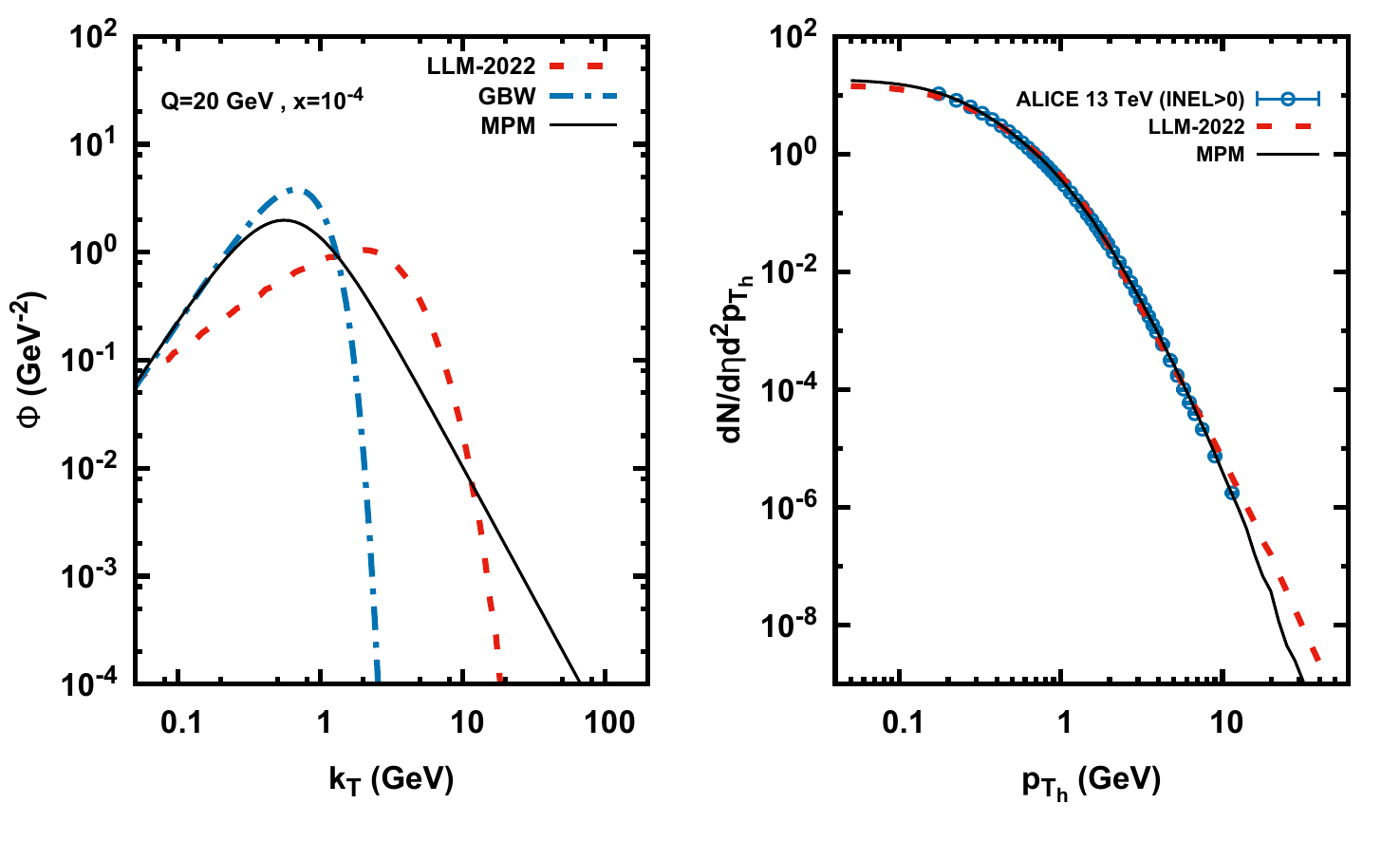}
\caption{Left plot: comparison of the corresponding UGDs: LLM-2022 (dashed line), GBW (dot-dashed line) and MPM (solid line). The LLM approach includes CCFM  evolution of the gluon function (see text).   Right plot: numerical results for MPM and LLM-2022 approaches compared to minimum bias data $\mathrm{INEL}>0$ for 13 TeV measured by ALICE Collaboration \cite{ALICE:2019dfi}.}\label{fig:comparingUGDs}
\end{figure*}

\section{Theoretical framework and main predictions}
\subsection{The $p_T$ spectrum in $k_T$-factorization}
\label{sec:modelA}
To carry out our phenomenological investigation we will employ the $k_T-$factorization formalism, where the cross section for particle production can be expressed in terms of the unintegrated gluon distribution (UGD), dependent on transverse momentum $k_T$. Different hard observables in $pp$, $pA$ and $AA$ collisions have been well described using this formalism \cite{Moriggi:2020qla,Moriggi:2022xbg,Moriggi:2023ahi}, in addition to diffractive processes \cite{Wusthoff:1999cr} as well as structure functions in interactions electron-proton ($ep$) and electron-nucleus ($eA$) \cite{Ivanov:2000cm,Luszczak:2022fkf,Bacchetta:2020vty,Lipatov:2013yra,Cisek:2014ala,Hentschinski:2021lsh,Lipatov:2022doa,deOliveira:2013oma,Lipatov:2023egx} (see also reviews on Refs. \cite{Angeles-Martinez:2015sea,Abdulov:2021ivr}). 
In all these approaches the fundamental element is the partonic dynamics represented by the color dipole scattering amplitude, whose Fourier transform in momentum space $k_T$ is directly connected with the gluon number of the target, $\phi(x,k_T) $. Different models have been proposed to this quantity for both protons and large nuclei \cite{Luszczak:2022fkf,GBWnovo,Kutak:2012rf,LT,Moriggi:2020qla,Ivanov:2000cm,ipsat,bCGC}, which include different considerations about partonic dynamics in relation to scaling, impact parameter ($\vec{b}$) and geometry dependence, high- $Q^2$ limit and so on.

Partons develop an anomalous diffusion-like dynamic in the two dimensional transverse momentum space $\vec{k}_T$, while its longitudinal dynamics is trivial. That can be described in the picture of QCD color dipoles \cite{Nikolaev:1990ja,Nikolaev:1991et,Mueller:1993rr,Mueller:1994jq}. One of the main features of this dynamic is the emergence of a saturation scale $Q_s(x)\sim x^{-1/3}$, which limits the growth of the gluon distribution at small Bjorken variable $x$. This behavior ends up being translated into observables, such as the $p_T$ spectrum of hadrons, where the cross section for hadronic production can be described by a universal function $f$, i.e. $d\sigma (p_T,\sqrt{s})/d^2p_T \sim f(\tau)$. Here, $\tau$ is the scaling variable which can be defined in the context of geometric scaling property of parton saturation approaches.

In the $k_T$ factorization formalism, the cross section for producing a gluon jet with transverse momentum $p_T$ results in the convolution of the non-integrated gluon distributions of the target and projectile,
\begin{eqnarray}
\label{eq:fatkt}
E\frac{d^3\sigma}{d\vec{p}^3}^{ab \rightarrow g+X} &=&\frac{\mathcal{A}}{p_T^2}\int d^2k_T\,\phi(x_a,k_T^2)\phi(x_b , q_T^2) , \\
 &= & f(\tau)  ,
\end{eqnarray}
which can be reduced to a universal function $f(\tau)$ due to the scaling in relation to the saturation scale $Q_s(x)$ characteristic of the parton saturation formalism. Namely, the scaling variable is given by $\tau = p_T^2/Q^2_s(x)$. Here, ${\mathcal{A}}=K\frac{2\alpha_s}{C_F}$ is the overall normalization and $\vec{q}_T = (\vec{p}_T-\vec{k}_T)$. The $K$ factor is a multiplicative constant as used in \cite{Moriggi:2020zbv}. The Casimir is $C_F = (N_c^2-1)/2N_c = 4/3$ and $\alpha_s$ is the strong coupling constant. The scaling curve for the case of gluon production in the range $\tau >1$ has the following form \cite{Moriggi:2020zbv},
\begin{eqnarray}
f(\tau)  = C\frac{\kappa}{\kappa - 1}\left[1-\frac{1+\kappa \tau}{(1+\tau)^{\kappa}}  \right]    \frac{1}{(1+\tau)^{1+\kappa}},
\end{eqnarray}
with $C$ being the overall normalization  and $\kappa = 1+\delta n$.

In order to illustrate how the UGD \cite{Moriggi:2020zbv} considered  here (Moriggi-Peccini-Machado - MPM -  parametrization) is compared to others unintegrated gluon functions in literature, in Fig. \ref{fig:comparingUGDs} (left plot) it is contrasted with the well known Golec-Biernat-Wüsthoff (GBW) parametrization \cite{GBW1,GBW2}, $\phi_{GBW}(x,k_T^2) = c_g(k_T^2/Q_s^2(x))e^{-k_T^2/Q_s^2(x)}$ and with the updated Lipatov-Lykasov-Malyshev (LLM) approach \cite{Lipatov:2022doa}, $\phi_{LLM}(x,k_T,Q^2)$. The GBW model and ours do not take into account the QCD evolution (the Catani-Ciafaloni-Fiorani-Marchesini (CCFM) evolution equations) of the gluon function, which depends also on the scale $Q^2$ of the problem. The LLM approach provides the UGD at low-$Q^2$ with a different analytical form from ours and  at large $Q^2$ the QCD evolution with CCFM equations is
calculated. The soft and hard hadron production have been computed and the model allows for a good data description (see more details in Ref. \cite{Lipatov:2022doa}). In the figure, the UGD is shown as a function of gluon transverse momentum, $k_T$, for fixed $x=10^{-4}$ and at hard scale $Q=20$ GeV. The saturation scales in GBW (dot-dashed line) and MPM (solid line) models are very close and scale independent, which is characterized by the transverse momentum peak of the distribution. The LLM approach (LLM-2022, dashed line) presents a higher saturation scale at this scale. In the right plot, the MPM and LLM approaches are compared in the description of the minimum bias data ($\mathrm{INEL}>0$) for 13 TeV at the LHC \cite{ALICE:2019dfi}. As expected, the QCD evolution starts to be important at $p_{T_h}\gsim 10$ GeV. It is worth mentioning that the parameters $K$ and $\langle z\rangle $ in the MPM model were obtained for the range $1<\tau_h<100$ with $\tau_h = p_{T_h}^2/Q_s^2(x_h)$.

Moving now to the multiplicity distributions, it is desirable constrict the scaling function in Eq. (\ref{eq:fatkt}) in terms of multiplicity degree of freedom, $f(\tau)\rightarrow f(\tau_i)$. Let us consider the variation of the saturation scale in each multiplicity class $i$ in relation to its minimum bias value, $X_i=Q_{s_i}(x)/Q_{s}(x)$, in the following way,
\begin{equation}
        \tau_i=\frac{Q^2}{\left[X_i Q_{s}(x)\right]^2}, 
\end{equation}
where the momentum scale involved in the hard interaction is given by $Q^2=p_T^2+m_j^2$, with $m_j$ being the mass of the produced gluon jet. The value for the jet mass used here is the same considered  in Ref. \cite{Moriggi:2020zbv}, $m_j=0.56$ GeV. The parton-hadron transition can be approximated by assuming that the hadron carries a fraction of momentum $\langle z \rangle $  of the parton. In the numerical calculations the value $\langle z \rangle \simeq 0.4$ will be used, which is determined from data on $pp$ collisions at the LHC \cite{Moriggi:2020zbv}. In addition, we should replace the gluon transverse momentum $p_{T}\rightarrow \frac{p_{T_h}}{\langle z \rangle}$, with $p_{T_h}$ being the hadron transverse momentum. Final state processes could destroy the spectrum scaling, but this does not happen, indicating that the role of these effects is secondary at least in $pp$ collisions.

We can assume that the variation in multiplicity is related to the variation in the collision impact parameter, which leads to an increase in the transverse area of proton overlap $\langle A_T \rangle $, up to a maximum area value associated with the total overlap $ \langle A_{T_{max}} \rangle $. The cross section can be expressed in a given multiplicity class scaling Eq. \eqref{eq:fatkt} in the form

\begin{equation}\label{eq:Nch}
    E\frac{d^3\sigma_{i}}{d\vec{p}^3}^{ab \rightarrow g+X}\sim \frac{\langle A_T \rangle}{\langle A_{T_{max}} \rangle} f(\tau_i).
\end{equation}
The multiplicity can be obtained from the cross section at a given multiplicity, $\sigma_i$, by assuming $\sigma_i/\sigma_{inel}=N_i/\langle N\rangle$. Here, $\sigma_{inel}$ is the inelastic cross section, $N_i$ and $\langle N \rangle $ are the number of produced particle in the multiplicity class and its average, respectively. The spectrum modification factor in relation to its minimum bias value $\langle d N/d^2p_{T_h}d\eta \rangle$ is defined as the ratio,
\begin{equation}\label{eq:ratio}
    R_i(p_{T_h})=\frac{d N_{i}/d^2p_{T_h}d\eta}{\langle d N/d^2p_{T_h}d\eta \rangle},
\end{equation}
which captures small variations in the slope of large $p_{T_h}$ at each centrality. Given these considerations we have only two parameters to be fitted in each multiplicity. They are the ratios between areas and saturation scale in relation to the reference value: i.e., $\frac{\langle A_T \rangle}{\langle A_{T_{max} } \rangle}$ and $X_i$.

We will not make an a priori estimate of these quantities, but in section \ref{sec:results} we will analyze the resulting behavior in terms of relevant indicators of multiplicity such as $\langle p_T \rangle_i$. Bulk properties can be inferred from integrated spectra:

\begin{eqnarray} \label{eq:interated}
\frac{dN_i}{d\eta}&=&\int d^2p_{T_h} \frac{dN_{i}}{d^2p_{T_h}d\eta}, \\
 \langle p_{T_h} \rangle_i &=& \int d^2p_{T_h} p_{T_h} \frac{dN_{i}/d^2p_{T_h}d\eta}{\langle dN/d\eta \rangle} .
\end{eqnarray}

The characterization of the gluon distribution, appearing in Eq. \eqref{eq:fatkt}, in order to produce the universal scaling function $f(\tau_i)$ and its particularities is given in the next section.

\subsection{Partonic Entropy}
\label{sec:modelB}

In the dipoles picture, the fundamental element used to describe the collision is the color dipole scattering amplitude $\mathcal{N}(x,\vec{r})$ in the dipole coordinate space ($\vec{r}$ is the transverse size of the color dipole). In the transverse momentum space its Fourier transform, $P (x,k_T)$, decodes target information by exchanging multiple gluons, making this object proof of the target's gluon distribution. Due to these multiple interactions, the scattering amplitude acquires a statistical character \cite{Levin:2005ff,Kharzeev:2017qzs} describing a diffusive process in the transverse momentum space in relation to the longitudinal momentum fraction $x$. The last quantity gives the inverse of diffusion time. Different models have been proposed for this object. In phenomenological terms, it was proposed \cite{Moriggi:2020zbv} that a good description of the $p_T$ spectrum of produced hadrons can be made considering the following distribution:
\begin{equation}\label{eq:MPM}
P^{MPM}(\delta n,x,k_T)=\frac{1+\delta n}{\pi Q^2_s(x)} \frac{1}{\left(1+k_T^2/Q_s^2(x)\right)^{2+\delta n}},
\end{equation}
where the parameter $\delta n $ takes into account deviation from the expected amplitude in leading order,  $\propto 1/k_T^4$. This distribution was initially proposed to describe the slope in the region of large $p_{T}$ in transverse momentum spectrum in $pp$ collisions. Moreover,  it can also provide a good description of the $p_T$ distribution in large systems like $pA$ and $AA$ interactions \cite{Moriggi:2020qla,Moriggi:2022xbg,Moriggi:2023ahi,Santos:2023yep,Ben:2022dmw}, as well as diffractive processes \cite{Peccini:2020jkj,Peccini:2021rbt,Cisek:2022yjj}, which prove a distinct kinematic region.
The number of gluons with a certain transverse momentum $k_T$ is given by 
\begin{equation}
\phi(x,k_T)= \frac{3}{4\pi^2 \alpha_s} k_T^2 P(x,k_T)
\end{equation}
considering a homogeneous target with  impact parameter dependence $\propto \Theta (R_p^2-b^2)$. We will investigate the dependence of the collision geometry of this object by letting the saturation scale depend on the multiplicity at each centrality.
The emergence of scaling in the variable $Q^2_s(x)\sim x^{-\lambda}$ is a remarkable property of the QCD in the high energy regime with ample experimental evidence \cite{Moriggi:2020zbv,McLerran:2014apa,Praszalowicz:2015dta,BUW,Praszalowicz:2013fsa,Osada:2019oor,Osada:2020zui,Stasto}. The successful phenomenological  GBW model \cite{GBW1,GBW2} describes this process by an exponential function on the scaling variable $k_T^2/Q_s^2$,
\begin{equation}\label{eq:GBW}
P^{GBW}(x,k_T)=\frac{1}{\pi Q^2_s(x)} \exp \left[-k_T^2/Q_s^2(x)\right].
\end{equation}

One can define an entropy indicator in the gluon diffusion process based on the Boltzmann - Gibbs (BG) statistics in the following way,
\begin{equation}\label{eq:BG}
    S^{BG}=-\int P(x,k_T) \log\left[ P(x,k_T)\right] d^2k_T,
\end{equation}
where all we need to describe partonic interactions is the quantity $Q_s(x)$. An immediate consequence of scaling in the variable $k_T^2/x^{-\lambda}$ is that the BG entropy is logarithmic in time $1/x$, and additive with respect to rapidity $Y=\log(1/x)$,
\begin{equation}
    S^{BG}=c_1+c_2\lambda\log(1/x),
\end{equation}
where $c_1,c_2$ are constants and $\lambda$ a function of $x$ and scale $Q^2$. Initially the partons are located in a small region around the saturation point, as time passes the proton becomes almost homogeneous in transverse momentum space and the entropy increases fast.
It is interesting to note that if $\lambda = \lambda (x,Q^2)$ as predicted in other models \cite{H1Lambda}, then the BG entropy may grow faster/slower and become non-additive. Therefore, models of this type when analyzed from the perspective of BG entropy, will not be extensive in relation to rapidity indicating a break in the scaling. One of the reasons for this break  and consequently entropy additivity is that gluons occupy a region $d$ in the transverse plane of the inverse order to their average transverse momentum $d \sim k_T^{-1} \sim Q_s(x)^{ -1}$. When a dependence  is introduced on the scale $Q \sim D^{-1}$ at which the proton structure is proved, one can suggest that the entropy becomes dependent on the resolution $Q^2/Q_s^2 \sim d^2/D^2$ in which  the system is seen. If $Q^2/Q_s^2\gg 1$ then we can resolve details of the partonic substructure and one should expect a decrease of entropy and we will have loss of information between different rapidity layers. 

In the distribution of Eq. \eqref{eq:MPM}, information about non-additivity is given by the parameter $\delta n$, given by
\begin{eqnarray}
\label{eq:pars}
  \delta n (\tau_i ) &=& 0.075 \, \tau_i^{0.188} ,\\
 Q_{s}^2(x)&=& \left( \frac{x_s}{x}\right) ^{1/3}. 
\end{eqnarray}

Now it should be noted that the distribution in Eq. \eqref{eq:MPM} can be obtained by maximizing the Tsallis entropy $S_q$,
\begin{equation}\label{eq:qentro}
    S_q=\int d^2k_T \frac{1-\left[ P(x,k_T) \right]^{q}}{q-1}. 
\end{equation}
if we identify,
\begin{eqnarray}
    q&=&\frac{3+\delta n}{2 + \delta n}, \\
    Q_s^{2\prime}&=&Q_s^2 (q-1),
\end{eqnarray}
and imposing the constrains \cite{Tsallis:1998ws}
\begin{eqnarray}
    \langle k_T^2\rangle_q = \frac{\int d^2k_T k_T^2 \left[P(k_T)\right]^q}{\int d^2k_T\left[P(k_T)\right]^q} &=& D_q, \\
    \int d^2k_T P(k_T) &=& 1.
\end{eqnarray}
By using the Lagrange method to find the optimizing distribution $P_{opt}\sim e_q^{-\beta k_T^2 }$, where $\beta$ is the Lagrange parameter, we can do the following identification, 
\begin{equation}
    \langle k_T^2\rangle_q = D_q = \beta^{-1}.
\end{equation}
Now, if we interpret the Lagrange parameter as the inverse of the temperature $\beta^{-1}=T$, using the scaling hypothesis:
\begin{equation}
    \langle k_T^2 (x) \rangle_q \sim \beta^{-1} (x_s/x)^{1/3} ,
\end{equation}
we have a generalization of Einstein relation for anomalous diffusion \cite{PhysRevE.60.2398,PhysRevLett.75.366}.

These two entropies $S^{BG}$ and $S_q$, Eqs.  \eqref{eq:BG} and \eqref{eq:qentro}, differ in non-additive character, except in the case $q=1$ where they coincide. Parton dynamics can be described in terms of the difference in rapidity between two layers in the parton cascade $\Delta Y = Y_a - Y_b = \log(x_a/x_b)$,
\begin{equation}
    S_q(Y_a+\Delta Y) \leq S_q(Y_a)+S_q(\Delta Y),
\end{equation}
which implies that there is a loss of information when comparing the two systems in relation to the BG case. Entropy in Eq. \eqref{eq:qentro} is non-additive, and non-additivity depends on the parameter $\delta n$, which will measure the loss of information when we compare the same situation described above to different values of this parameter.

In general terms, we can argue that if the steady state of the distribution has the form \eqref{eq:MPM}, a natural choice for entropy would be \eqref{eq:qentro}. Considering that in the high energy regime $Q_s(x)^2/Q^2 \rightarrow \infty$ we expect a point particle scattering $\phi(k_T)_{LO}\sim k_{T}^{- 2}$ characterized by $q_{\infty}=3/2$. In practice, $q$ is always close to 3/2, which justifies the approximation $q\simeq 3/2+ \delta n(\tau_i) /4$. The variation of $q$ as a function of the scale tested in the system was modeled as a power in the form \eqref{eq:pars}. Such behavior of the entropic index is justified in analogy to other similar physical systems that exhibit this behavior \cite{Tsallis_2002},
\begin{equation}\label{eq:q-tau}
    q_{\infty}-q \simeq  \tau_i^{0.188}.
\end{equation}
and the value of the entropic index must increase with $Q^2$. The resulting entropy is given by,
\begin{equation}\label{eq:entroQ32}
    S_{q}(Q^2,Q_s^2)=\frac{1}{q-1}-\left(\frac{2-q}{q-1}\right)^q (\pi Q_{s_i}^2)^{1-q}
\end{equation} 
This entropy can be expressed in terms of the partonic kinematic variables $x,Q^2$ or expressed in terms of the spectrum variables, $p_{T_h}, \sqrt{s}$ in each multiplicity. As $q>1$ the entropy grows more slowly than $\log(1/x)$ eventually saturating.

\begin{table*}[t]
\begin{tabular}{|l|l|l|l|l|l|l|l|l|l|} 
\hline
                     & Multiplicity Class & $X_i$   & $A_T/A_{T_{max}}$ & $\chi^2/dof$ &                    & Multiplicity Class & $X_i$ & $A_T/A_{T_{max}}$ & $\chi^2/dof$ \\ \hline
$\sqrt{s}=5.02 $ TeV & I                  & -       & -                & -                                        & $\sqrt{s}=13 $ TeV & I                  & 1.765 & 0.777            & 0.86     \\
                     & II                 & 1.588  & 0.775            & 0.25                                    &                    & II                 & 1.567 & 0.849            & 0.68     \\
                     & III                & 1.489  & 0.782            & 0.17                                     &                    & III                & 1.486 & 0.837            & 0.55     \\
                     & IV                 & 1.407  & 0.782            & 0.14                                     &                    & IV                 & 1.402 & 0.843            & 0.53     \\
                     & V                  & 1.326  & 0.772            & 0.14                                     &                    & V                  & 1.323 & 0.831            & 0.48     \\
                     & VI                 & 1.235  & 0.758            & 0.16                                     &                    & VI                 & 1.240 & 0.809            & 0.42     \\
                     & VII                & 1.133  & 0.741            & 0.16                                     &                    & VII                & 1.146 & 0.778            & 0.28  \\
                     & VIII               & 1.012  & 0.714            & 0.20                                     &                    & VIII               & 1.036 & 0.731            & 0.20    \\
                     & IX                 & 0.845 & 0.698            & 0.27                                     &                    & IX                 & 0.882 & 0.680            & 0.14     \\
                     & X                  & 0.603 & 0.582            & 0.42                                     &                    & X                  & 0.637 & 0.498            & 0.06     \\ \hline
\end{tabular} 
\caption{Fitted parameters $X_i$ and  $A_T/A_{T_{max}}$ from experimental data \cite{ALICE:2019dfi} at the energies 5.02 TeV and 13 TeV in each multiplicity class.} \label{tab:pars}
\end{table*}

Another way of looking at the difference between exponential distribution and power law is through the superstatistics framework of Beck and Cohen \cite{Beck_2001} where the Tsallis distribution is obtained from the gamma fluctuations of the inverse of the saturation scale $\beta = 1/Q^2_s(x)$. In our case one has,
\begin{equation}\label{eq:superst}
    \int d\beta P^{GBW}(k_T,\beta) g(n,\beta,\beta_0) = P^{MPM}(k_T,\beta_0),
\end{equation}
where $g(n,\beta,\beta_0)$ is the gamma distribution. It is interesting to note that gamma distributions are needed in models like 
 those shown in Refs. \cite{Moreland:2018gsh,Carzon:2021tif} to generate the necessary multiplicity fluctuation in $pp$ and $AA$ collisions.

The growth of $q$ with $Q^2$ is consistent with the statistical argument of Ref. \cite{Beck_2001} in the analysis of turbulent flow, where the authors argue that the variance in the fluctuations of $\beta$ are smaller if taken at a larger distance scale, so $q$ must grow with $Q^2$. It is noteworthy that the parameter $q=3/2$ expected in the QCD parton picture is the same as that described by those authors in the context of turbulent flow on small scales.

Finally, the use of the proposed entropy offers a simple and economical way in terms of number of parameters to describe experimental data of the $p_{T}$ spectra and makes clear the partonic dynamics in a given collision process. In next section we explore the connection between the multiplicity
of produced hadrons and partonic dynamics at high energies by using the indicator of partonic entropy associated and investigate the  relationship between the growth of entropy, the area of interaction and the final multiplicity of charged hadrons.

\begin{figure*}[t]
\includegraphics[width=0.8\textwidth]{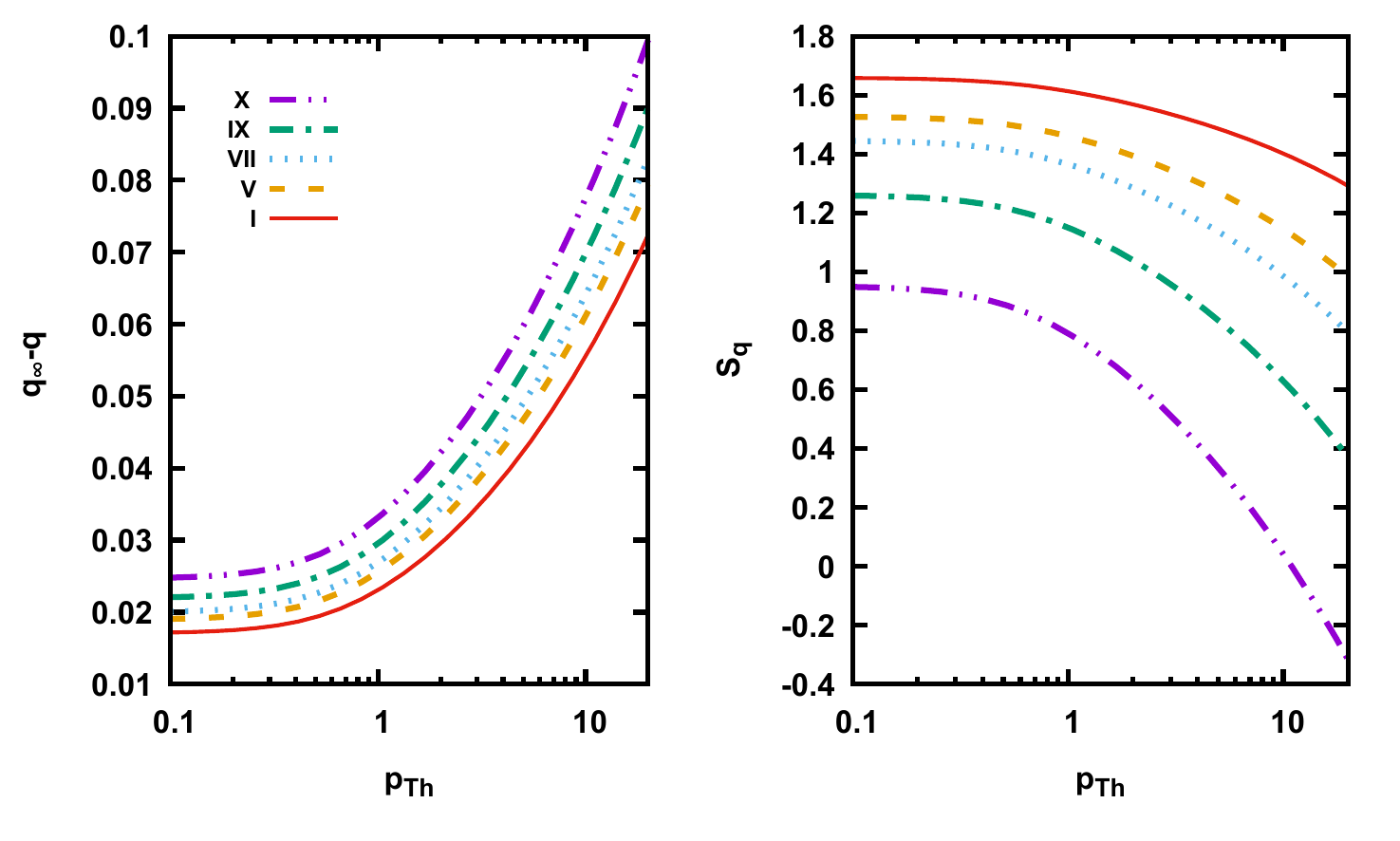}
\caption{Left plot: evolution of the entropic index with the hadron transverse momentum, given by Eq. \ref{eq:q-tau}. Right plot: partonic entropy $S_q$ given by Eq. \ref{eq:qentro} for different multiplicities. In both plots representative multiplicity classes are considered (I,V,VII,IX, X)}\label{fig:qent}
\end{figure*}

\section{Results and discussions}
\label{sec:results}

\begin{figure*}[t]
\includegraphics[width=0.8\textwidth]{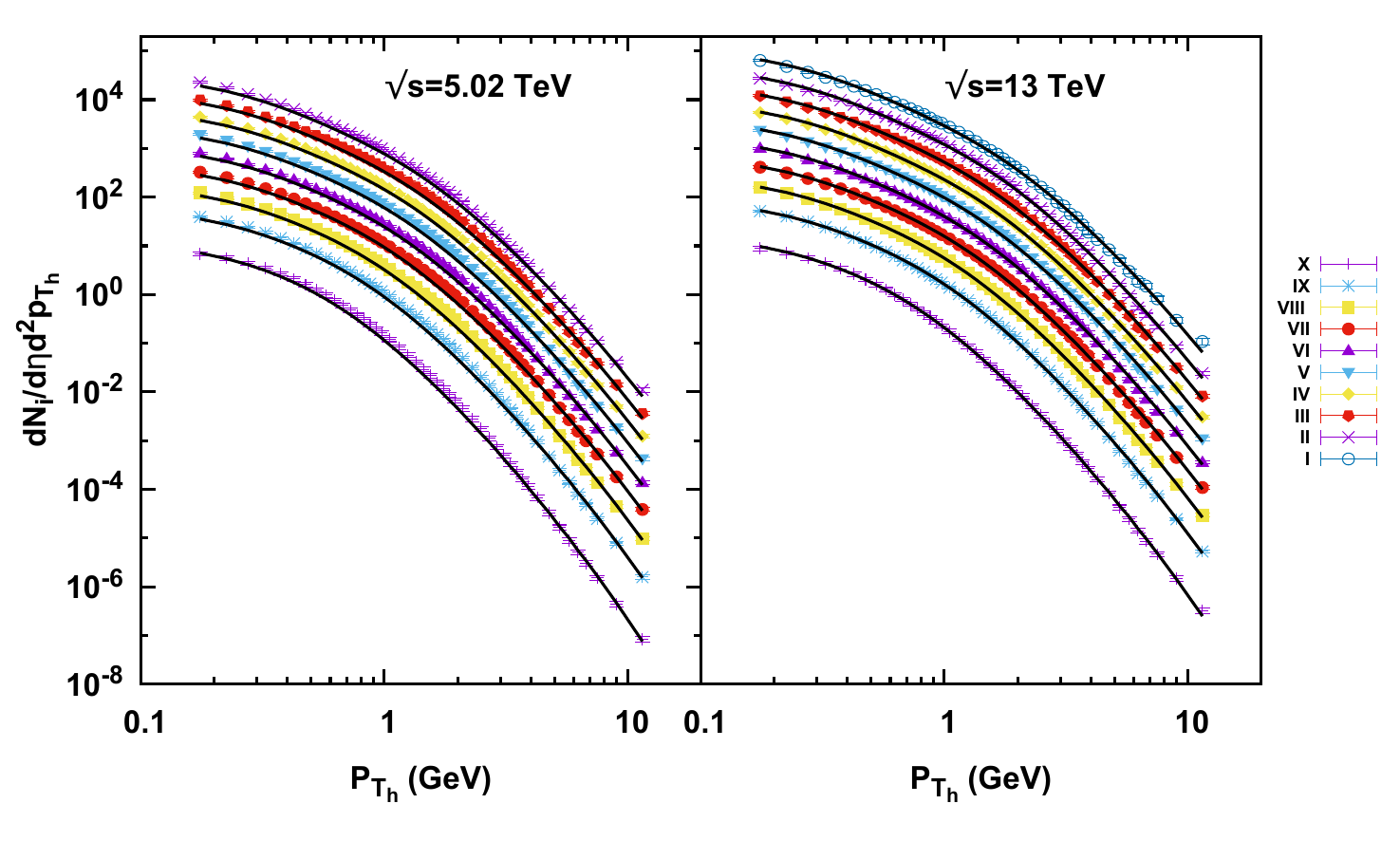}
\caption{The $p_{T_h}$ spectrum for different multiplicities, comparison of ALICE data\cite{ALICE:2019dfi} with the result of adjusting the saturation scale given by Eq. \eqref{eq:Nch}. The data is multiplied by a factor $2^i$ for better visualization.}\label{fig:espectro}
\end{figure*}

First, we have addressed the issue of the centrality classes. For each ALICE multiplicity class \cite{ALICE:2019dfi} (energies  5.02 and 13 TeV ), a value was adjusted for the parameter $X_i$, which measures the deviation in relation to the average saturation scale $X_i=Q_{s_i}(x)/Q_{s}(x)$, in addition to the transverse area ratio $\langle A_{T} \rangle/ \langle A_{T_{max}} \rangle$. We consider the range $p_{T_h} < 10 $ GeV, where scaling is observed with good precision. The resulting parameters are presented in Table \ref{tab:pars} for the two collision energies. The multiplicity classes are labeled by ten identifiers (index I to  X). Although the saturation scale grows with each multiplicity, the interaction area saturates at a limit close to $\langle A_{T_{max}} \rangle$ characterizing the total overlap of the protons. This trend is expected, as it has been observed in \cite{Bzdak:2013zma,McLerran:2014apa}.

The resulting $p_T$ spectra given by Eq. \eqref{eq:Nch} are presented in Fig. \ref{fig:espectro} for each multiplicity class at energies of $\sqrt{s}=5.02$ TeV and $\sqrt{s}=13$ TeV and compared to data from ALICE \cite{ALICE:2019dfi}. In order to convert multiplicity to cross section the inelastic cross section used was $\sigma_{inel}(\sqrt{s}=5.02 )/\sigma_{inel}(\sqrt{s}=13) = 0.87 $ \cite{Loizides:2017ack}.
Considering the large momentum, $p_{T_h}\approx 10$ GeV, we observe that the slope is smaller for events of high multiplicity (like class I) than for low multiplicity (for instance in class X) as a consequence of the increase in partonic entropy \eqref{eq:qentro} in events of high multiplicity. The entropy for these situations is presented in Fig. \ref{fig:qent}, along with the entropic index $q$ associated with the same $p_{T_h}$ region. As entropy increases, so does multiplicity, but this growth is faster for interactions with large transferred momentum $Q^2\sim p^2_{T_h}$. In the limit where $p_{T_h}/Q_s(x) \rightarrow 0$ the entropy would be flat, establishing a limit for particle production in this kinematic region.

\begin{figure*}[t]
\includegraphics[width=0.8\textwidth]{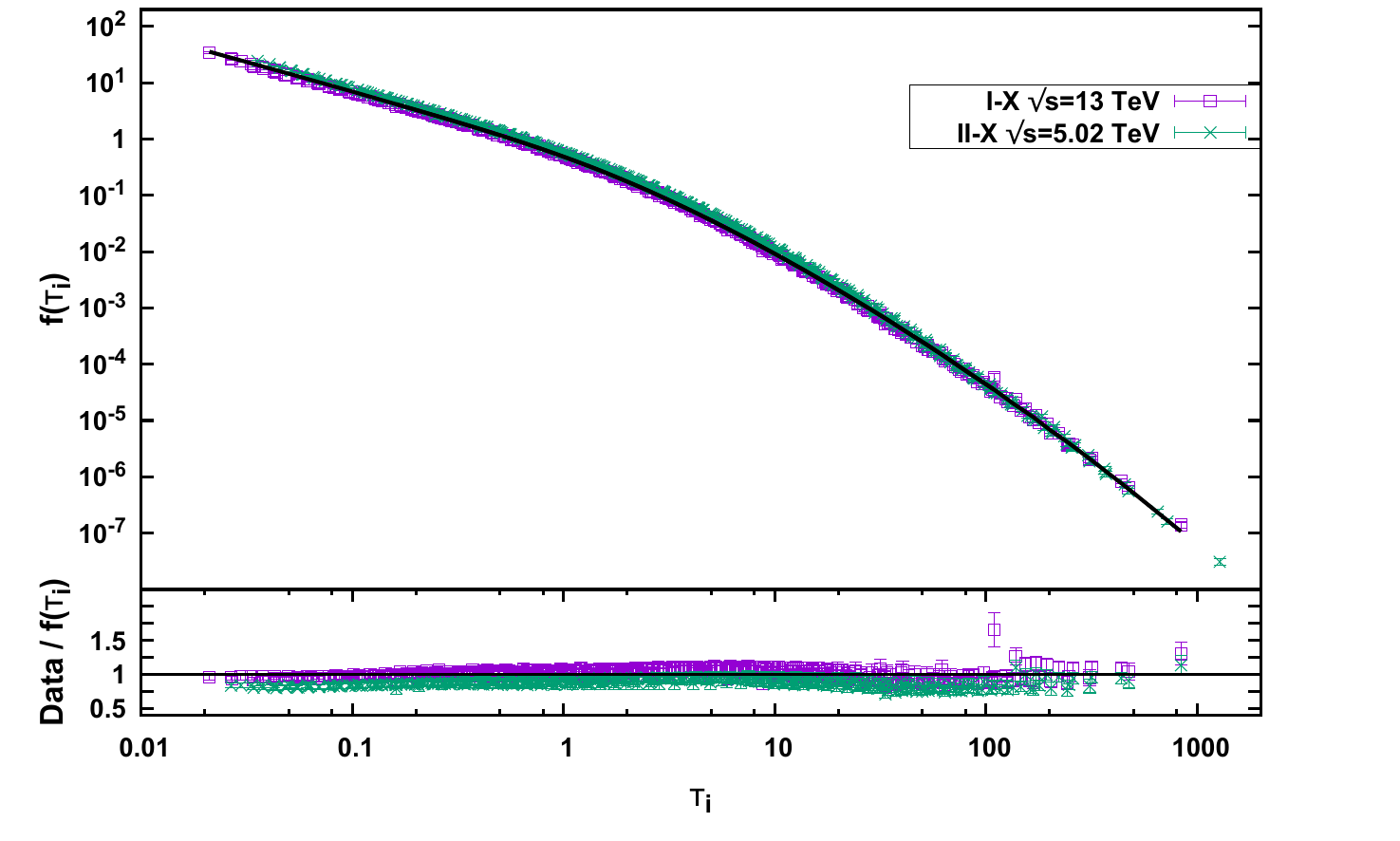}
\caption{Scaling function  obtained by the relation \eqref{eq:Nch}. Comparison is done between scaling behavior obtained from data at different multiplicities and the scaling curve calculated in the $k_T$ factorization formalism.}\label{fig:scaling}
\end{figure*}

The scaling in the universal function $f(\tau_i)$ is evident in the spectrum shown Fig. \ref{fig:scaling} at different energies as  the data/theory error is very close to 1.  A relevant deviation is only seen for the region of $\tau_i\sim 10^3$ where the validity of the model is in its limit of application.  This sort of scaling in $pp$ collisions has been verified also in Ref. \cite{Osada:2019oor},  where the multiplicity dependence has been embedded on the saturation momentum  within the geometrical scaling approach. The corresponding scaling is assumed in both semi-inclusive and inclusive distributions.

\begin{figure*}[t]
\includegraphics[width=0.8\textwidth]{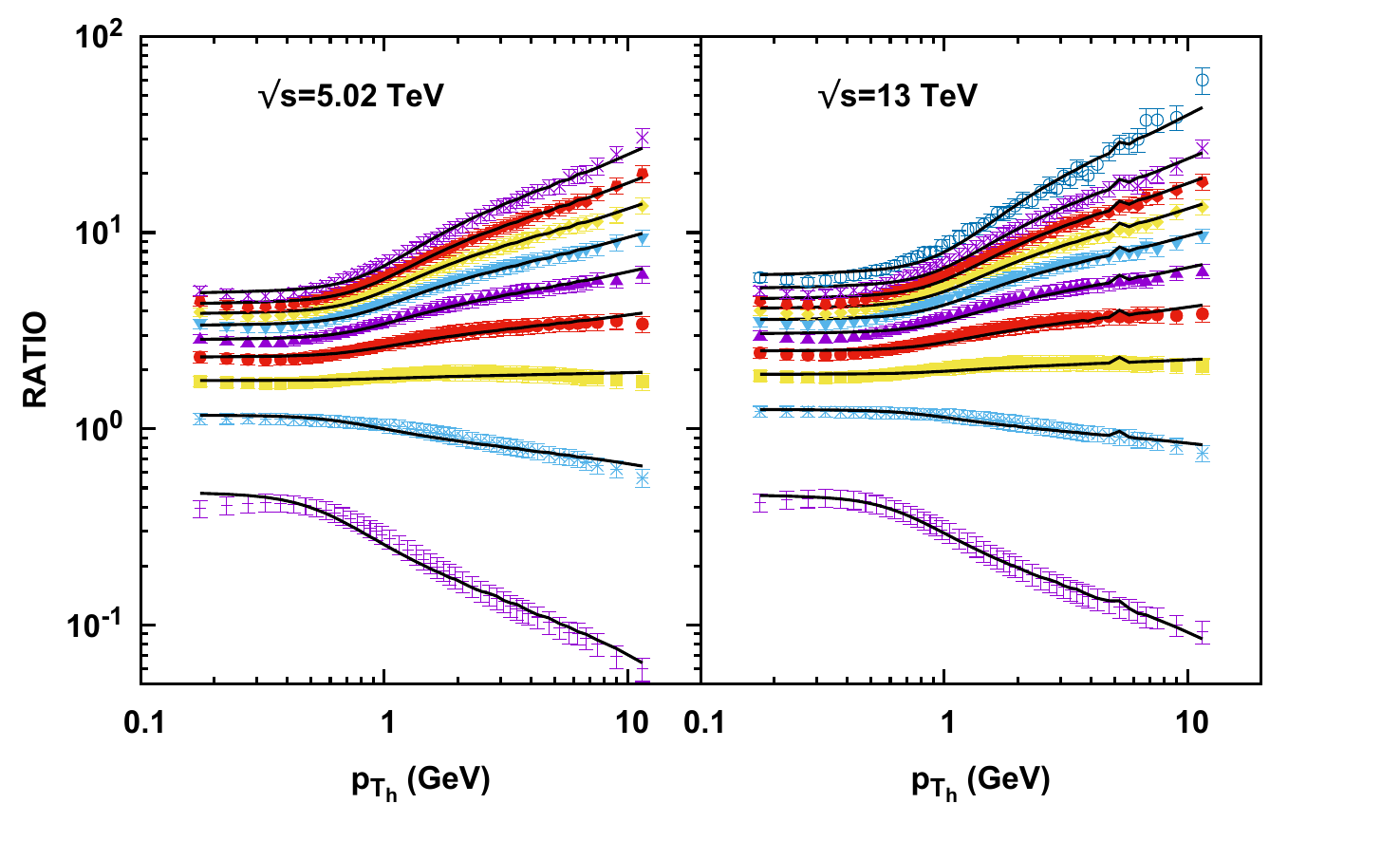}
\caption{Ratio between different multiplicity classes and $\langle \frac{dN}{d^2p_{T_h}d\eta}\rangle$, calculated by Eq.\eqref{eq:ratio}, as a function of hadron transverse momentum $p_{T_h}$. Results are shown for the energies of 5.02 TeV (left plot) and 13 TeV (right plot) and the theoretical prediction represented by solid lines.}\label{fig:ratio}
\end{figure*}

\begin{figure*}[t]
\includegraphics[width=0.8\textwidth]{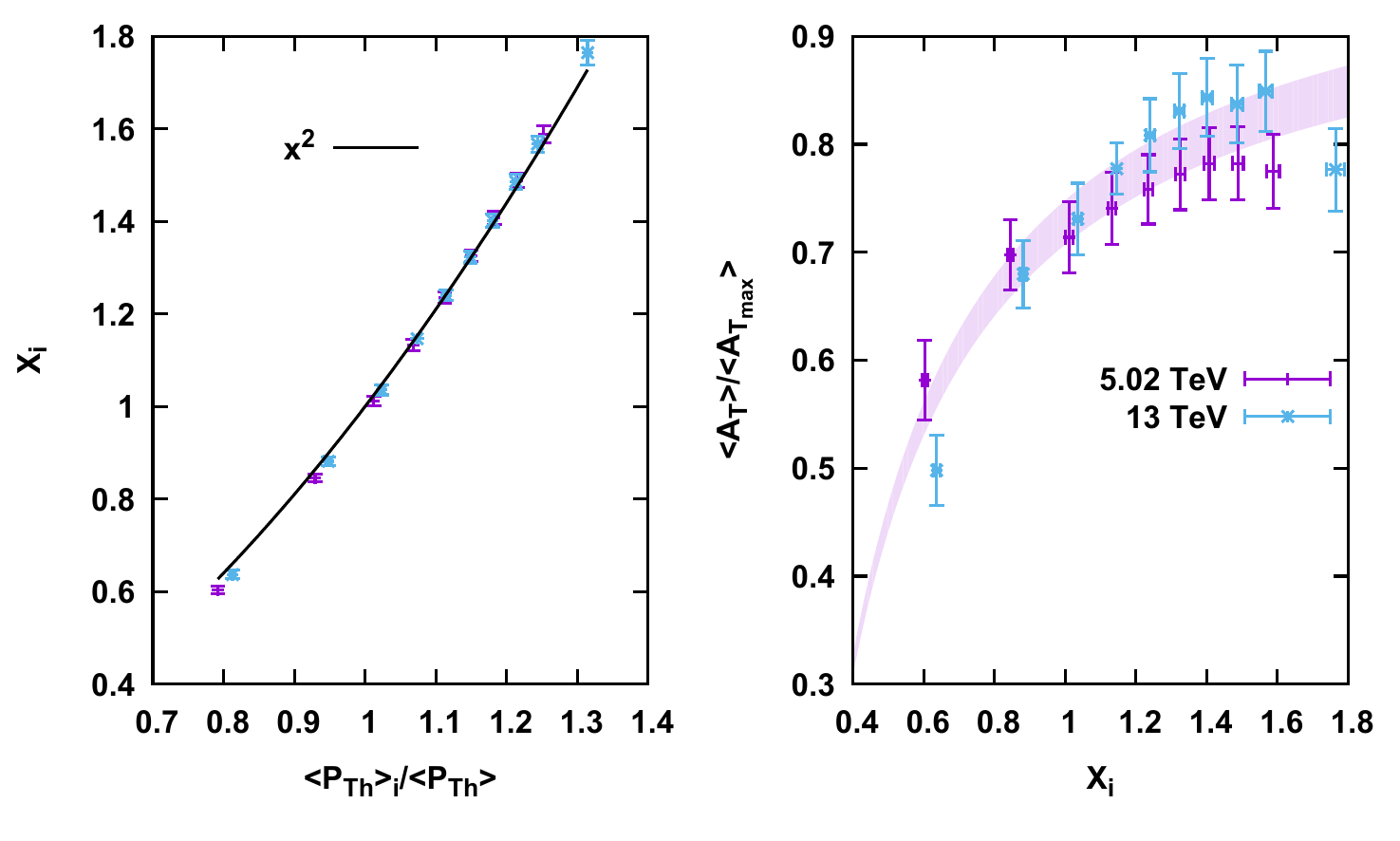}
\caption{Left plot: The values $X_i=Q_{s_i}/Q_{s}$ for saturation scale of each centrality class (see Table \ref{tab:pars}) as a function of the average transverse momentum ratio, $u=\langle p_{T_h} \rangle_i/\langle p_{T_h}\rangle$. The solid line corresponds to a quadratic behavior, $X_i=u^2$.  Right plot: Average interaction area normalized by its maximum value  as a function of $X_i=Q_{s_i}/Q_{s}$ or $ u^2$. The band corresponds to a non-linear fit based in the form shown in Eq. \eqref{eq:s32} by using the proportionality proposed in Eq. \eqref{eq:prop-AT-S}. }
\label{fig:ptqs}
\end{figure*}

The ratio between the multiplicities classes and $\langle dN/d^2p_{T_h}d\eta\rangle$, defined in Eq. \eqref{eq:ratio}, is presented in Fig. \ref{fig:ratio} for the energies of 5.02 TeV (left plot) and 13 TeV (right plot) as a function of hadron transverse momentum, $p_{T_h}$. This observable is interesting because the ratio is sensitive to small variations in the slope on $p_{T_h}$. We can see that our model provides a good description of the data for all multiplicities. It is interesting to note that the slope of the spectrum given by $\delta n$ in the distribution of Eq. \eqref{eq:MPM} determines the growth of the ratio in the region of large $p_{T_h}$. It is not a priori expected that parameter $\delta n$ of the Eq. \eqref{eq:pars} could provide the appropriate slope for each multiplicity just by rescaling $Q_s \rightarrow X_i Q_s$ without any extra parameters. As shown in Fig. \ref{fig:ptqs} (left plot), where $X_i =(\langle p_{T_h}\rangle_i/\langle p_{T_h}\rangle)^2$, the relationship  established is that the spectrum slope in each multiplicity class can be derived from its minimum bias multiplicity  by just rescaling $Q_s \rightarrow \langle p_{T_h}\rangle^2 $ in the UGD power index, $\delta n$. 

\begin{figure}[t]
\includegraphics[width=1.\linewidth]{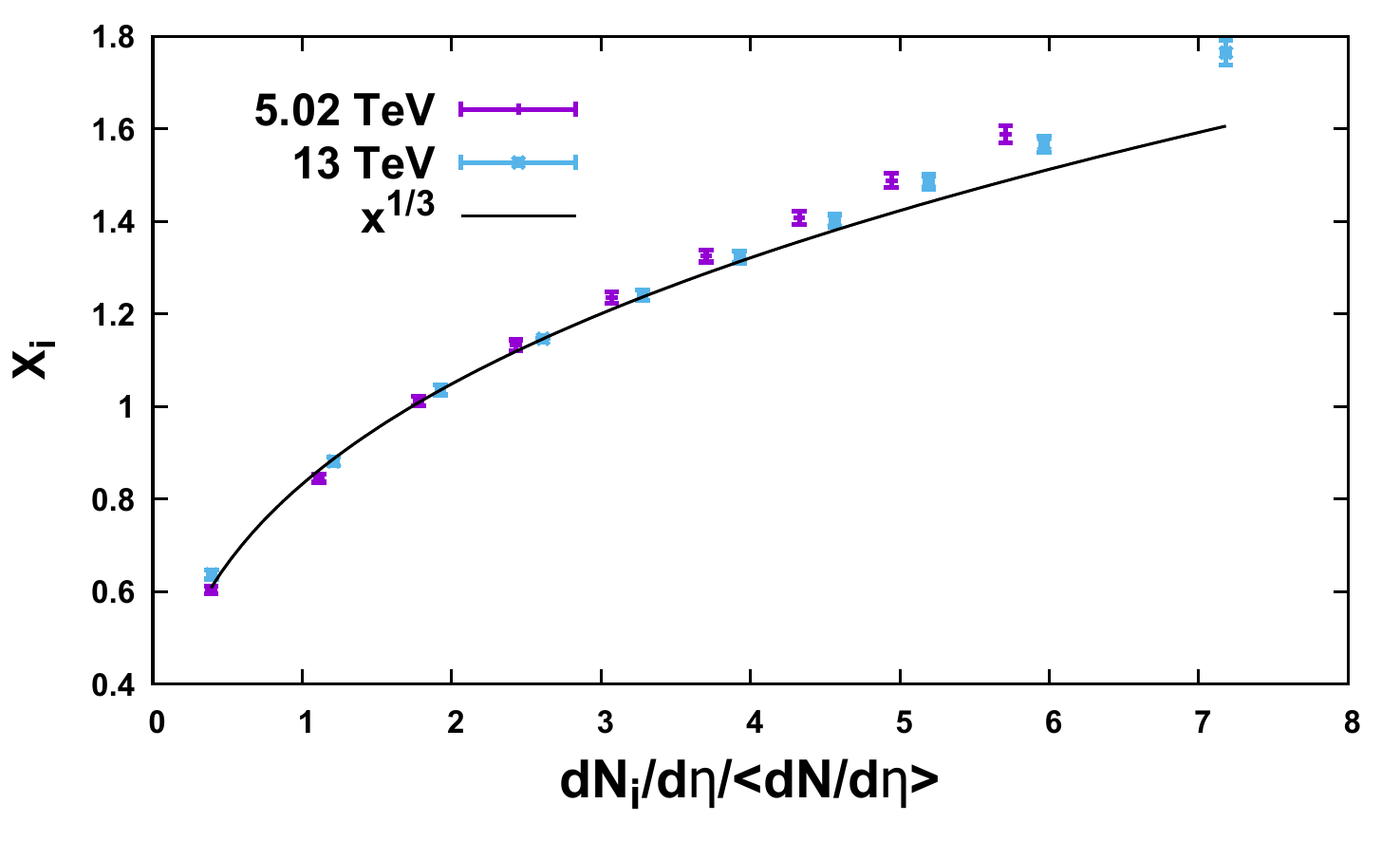}
\caption{Saturation scale as a function of multiplicity expressed in terms of the relation between the quantity $X_i$ and $dN_i/d\eta$. Solid line corresponds to a power-like fit to the relation.}
\label{fig:NCHQS}.
\end{figure}

Concerning the scaling of $Q_s$ on multiplicity in the context of parton saturation approaches it is expected that for high multiplicity events, based on the Local Parton Hadron Duality (LPHD), the density of gluon grows as a function of multiplicity \cite{Kharzeev:2001yq,Kharzeev:2007zt,Dumitru:2011wq,Kovchegov:2000hz,Lappi:2011gu}. This leads by consequence to multiplicity dependence of the saturation scale. The integrated spectra at given energy \eqref{eq:interated} under the scaling can be expressed as 

\begin{equation}
    \frac{dN_{i}}{d\eta} \sim \frac{\langle A_T\rangle}{\langle A_{T_{max}}\rangle} X_i^2.
\end{equation}
The specific shape of the overlap area dependence and the saturation scale with multiplicity can give us important information about partonic dynamics.
In our model, the saturation scale growth due the multiplicity $d N_i/d\eta$ can be approximated by
\begin{equation}
    X_i\sim \left(\frac{d N_i/d\eta}{\langle d N/d\eta \rangle} \right)^{1/3},
\end{equation}
 which is shown in Fig. \ref{fig:NCHQS}. This behavior is different from the one obtained in Ref. \cite{Osada:2019oor} where $X_i\sim c_1+c_2 (dN_i/d\eta)^{1/6}$ or the linear behavior used in \cite{Levin:2010dw,Levin:2019fvb}.

Now, in order to understand the relation between area of interaction, multiplicity and entropy, let us calculate the entropy relative to the distribution on $1/Q_{s}^2(x)$ approximated by $q=3/2$. This provides the following result:
\begin{equation}\label{eq:s32}
    S_{3/2}(X_i)=2\left( 1- \frac{1}{\left(\pi X_i^2\right)^{1/2}}\right).
\end{equation}

If we assume that entropy is extensive with respect to the area of interaction, then we come to the conclusion that
\begin{equation}
S_{3/2}(X_i) \sim \langle A_T \rangle \sim (dN_{i}/d\eta)^{1/3},
\label{eq:prop-AT-S}
\end{equation}
with $\langle A_T \rangle$ (and $S_{3/2}(X_i)$)  reaching a saturation in a maximum value at large multiplicities.

In Figure  \ref{fig:ptqs} (right plot)  the average interaction area is presented as a function of the saturation scale ratio, $X_i$. By using the proportionality between the average interaction area and the entropy, Eq. \ref{eq:prop-AT-S}, in Fig. \ref{fig:ptqs} the results are shown for a fit in  the form  of Eq. \eqref{eq:s32}. Namely, the relation $\frac{\langle A_T \rangle}{\langle A_{T_{max}} \rangle}=\xi (1-a/X_i)$ has been used, where $\xi$ is a proportionality constant.  The value found for the parameter $a=0.67 \pm 0.06$ is close to $1/\sqrt( \pi)=0.56$ appearing in Eq.  \eqref{eq:s32}.  The dependence of the interaction area on multiplicity is investigated in Refs. \cite{Bzdak:2013zma,McLerran:2014apa,Osada:2019oor}, where it is argued that the interaction  area has a natural dependence on multiplicity in the form $\langle A_T \rangle \sim (dN_{i}/d\eta)^{2/3}$ (scales with the volume $R^3$), until its saturation at a certain limit for high multiplicities. If we assume the scaling of the transverse area with partonic entropy we have a different result, that is $\langle A_T \rangle \sim (dN_{i}/d\eta)^{1/3}$.

Finally, putting the present work in context in Ref. \cite{Herrera:2024zjy} the  normalized transverse momentum distributions of produced hadrons,  have been used to compute the Boltzmann-Gibbs entropy. The heat capacity is also determined from the entropy. It has been considered statistics of three different fitting functions: thermal, confluent hypergeometric 
and the Hagedorn distribution. Minimum bias data in $pp$ collisions at RHIC and LHC have been considered and it was shown that the BG entropy of the final state increases with the collision energy. In Ref. \cite{Caputa:2024xkp}, it was proposed a connection between the framework of the evolution of states with dynamical $SL(2,R)$ symmetry on the context of the Krylov basis and the evolution of QCD color dipoles in Mueller dipole cascade framework. The latter has been used to define a parton (mostly gluons at large rapidities) entanglement entropy, $S_{E}$ \cite{Kharzeev:2017qzs}. In the simplest case of (1+1) dimension, one obtains $S_E=\ln [xg(x)]$, where $xg(x)$ is the integrated gluon distribution.  At high energies the hadron becomes a maximally entangled state and the multiplicity distributions in deep inelastic scattering and hadron-hadron scattering can be deduced from the QCD parton cascade \cite{Levin:2023mwl,Gotsman:2020bjc}. Work 
 in \cite{Caputa:2024xkp} connects the K-complexity to the number
 of color dipoles in the parton cascade and the K-entropy to their $S_E$. On the other hand, a \textit{dynamical entropy} for dense QCD states was proposed in Ref.  \cite{Peschanski:2012cw}, which is based on statistical physics tools for far-from-equilibrium processes. The numerical analysis by using realistic gluon UGDs has been done in Ref. \cite{Ramos:2022gia}. This entropy is written as an overlap functional between the gluon distribution at different total rapidities $Y$ and saturation radius, $R_s(Y)=1/Q_s(Y)$, where $Q_s(Y)$ is the saturation scale. In the weak coupling regime the dynamical entropy characterizes the change of the color correlation length $R_s(Y_0)\rightarrow R_s(Y)$, mirroring  the rapidity evolution $Y_0\rightarrow Y$ of a dense gluon state. The entropy functional $\Sigma^{Y_0\rightarrow Y}$ is defined in terms of the gluon transverse momenta probability distribution. In some aspects, the analysis presented here is more directly connected to this dynamical entropy. 
 
 \section{SUMMARY AND CONCLUSIONS} \label{sec:conclusion}
In this work we present a description of the $p_T$ spectra of charged hadrons produced in high-energy collisions taking into account their dependence on multiplicity. In order to do so, we interpret the distribution of gluons in terms of their entropy and show how characteristics of the spectrum can be well described in terms of this quantity plus the partonic saturation scale. We show that multiplicity data exhibit scaling in relation to the saturation scale. An important consequence of our formulation is that although the partonic entropy is non-extensive with respect to the rapidity as expected from the BG statistics, when we compare the area of interaction of the protons it appears that the partonic entropy grows with the interaction area. This may suggest that the dependence on geometric aspects is fundamental for partonic dynamics, which is usually not included in the evolution of the rapidity in the distributions. Finally, we note that the correct slope of the spectrum under the rescaling of the UGD power parameter, as well as the economic number of parameters and the linear dependence of entropy on the partonic interaction area are the main results of this work.

\section*{Acknowledgments}

This study was financed in part by the Coordena\c{c}\~ao de Aperfei\c{c}oamento de Pessoal de N\'{\i}vel Superior - Brasil (CAPES) - Finance Code 001. MVTM
acknowledges funding from the Brazilian agency Conselho Nacional de Desenvolvimento Cient\'ifico e Tecnol\'ogico (CNPq) with grant CNPq/303075/2022-8.

\bibliographystyle{h-physrev}
\bibliography{referencias_pp}

\end{document}